\documentclass[pre,twocolumn,showpacs,preprintnumbers,amsmath,amssymb]{revtex4}

\usepackage{graphicx}
\usepackage{dcolumn}
\usepackage{bm}

\begin{document}


\title{Physics of Psychophysics: Stevens and Weber-Fechner laws are
transfer functions of excitable media}

\author{Mauro Copelli}%
\email{mcopelli@df.ufpe.br}
\affiliation{Instituto de F\'\i sica, Universidade Federal Fluminense, 
Av. Litor\^anea, s/n, 24210-340, Niter\'oi, RJ, Brazil}
\author{Ant\^onio C. Roque}
\email{antonior@neuron.ffclrp.usp.br}
\author{Rodrigo F. Oliveira}
\email{rodrigo@neuron.ffclrp.usp.br}
\author{Osame Kinouchi}
\email{osame@dfm.ffclrp.usp.br}
\affiliation{Departamento de F\'{\i}sica e Matem\'atica, 
Faculdade de Filosofia, Ci\^encias e Letras de Ribeir\~ao Preto, 
Universidade de S\~ao Paulo, 
Av. dos Bandeirantes 3900,  14040-901, Ribeir\~ao Preto, SP, Brazil}

\begin{abstract}
Sensory arrays made of coupled excitable elements can improve both
their input sensitivity and dynamic range due to collective non-linear
wave properties. This mechanism is studied in a neural network of
electrically coupled (e.g. via gap junctions) elements subject to a
Poisson signal process. The network response interpolates between a
Weber-Fechner logarithmic law and a Stevens power law depending on
the relative refractory period of the cell. Therefore, these non-linear
transformations of the input level could be performed in the sensory periphery
simply due to a basic property: the transfer function of excitable media.

\end{abstract}


\maketitle

\bigskip

A very common trade-off problem found in the biology of sensory
mechanisms (and sensor devices in general) is the competition between
two desirable goals: high sensitivity (the system ideally should be
able to detect even single signal events) and large dynamic range (the
system should not saturate over various orders of magnitude of input
intensity).  In physiology, for example, broad dynamic ranges are
related to well known psychophysical laws \cite{Stevens,BBS}: the response $R$
of the sensory system may be proportional not to the input level $I$ but
to its logarithm, $R \propto \ln I$ (Weber-Fechner Law) or to a power of it,
$R \propto I^\alpha, \:(\alpha<1)$ (Stevens Law).

Most of the attempts to explain these psychophysics laws consist basically in
top-down approaches trying to show that they could
be derived from some optimization criterium for information processing \cite{BBS,Chater}.
In this work we use a botton-up, statistical mechanics approach, showing
how these laws emerge from a microscopic level. Indeed, they are generic transfer
functions of excitable media subjected to external (Poisson) input. Of course,
this does not explain ``why'' these laws have been adopted by Biology (some optimization
criterium may be relevant here), but explains why
Biology uses excitable media to implement them.

Receptor cells of sensory systems are electrically
coupled via gap junctions \cite{Dorries,ZM}. However, the 
functional roles of this electrical coupling are largely unknown. Here we report a simple
mechanism that could increase at the same time the sensitivity and the
dynamic range of a sensory epithelium by using only this electrical
coupling. The resulting effect is to transform the individual
linear-saturating curves of individual cells into a collective
Weber-Fechner like logarithmic response curve with high sensitivity to
single events and large dynamic range. We also observe a change to
power law behavior (Stevens Law) if relative
refractory periods are introduced in the model.

Although the phenomenon discusssed in this
work could be illustrated at different modeling levels \cite{Kinouchi},
we have chosen here to work with the
simplest elements: cellular automata (CA). The simplicity of the model
supports our case that the mechanisms underlying the described
phenomena are very general. To confirm this picture, we also present
preliminary results for neurons modeled by the Hodgkin-Huxley
equations. 

The $n$-state CA model is an excitable element containing two
ingredients: 1) a cell spikes only if stimulated while in its resting
state and 2) after a spike, a refractory period takes place, during
which no further spikes occur, until the cell returns to its resting
state. Denoting the state of the $i$-th cell at time $t$ by $x_i(t)\in
\{0,1,\ldots,n-1\}$, the dynamics of the proposed CA can be simply
described by the following rules:
\begin{enumerate}
\item If $x_i(t)=0$, then $x_i(t+1)= h_i(t)$, where $h_i\in \{ 0,1
\}$.
\item If $x_i(t)\neq 0$, then $x_i(t+1)=[x_i(t)+1]\mod n$. 
\end{enumerate}
Interpretation of the above rules is straightforward: a cell only
responds to stimuli in its resting state ($x_i=0$). If there is no
stimulus ($h_i=0$), it remains unchanged. In case of stimulus
($h_i=1$), it responds by spiking ($x_i=1$) and then remaining
insensitive to further stimuli during $n-2$ time steps ($x_i \in
\{2,\ldots,n-1\}$). 

In what follows, we assume that the external input signal $I_i(t)$
arriving on cell $i$ at time $t$ is modeled by a Poisson process of
supra-threshold events of stereotyped unit amplitude: $I_i(t)= \sum_n
\delta\left( t,t^{(i)}_n\right)$ where $\delta(a,b)$ is the Kronecker
delta and the time intervals $t^{(i)}_{n+1}-t^{(i)}_{n}$ are
distributed exponentially with average (input rate) $r$, measured in
events per second. For uncoupled cells, we have then simply $h_i(t) =
\delta\left( I_i(t),1\right)$.

\begin{figure}
\begin{center}
\includegraphics[width=0.8\columnwidth]{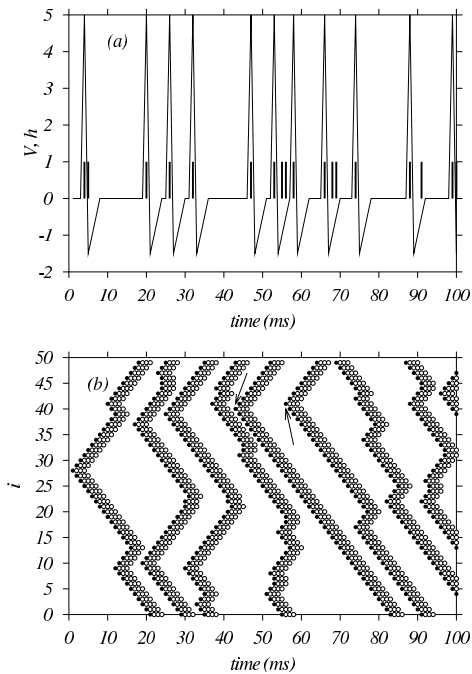}
\caption{\label{fig:dyn}Time evolution for $n=5$: (a) $V(x(t))$ for a
single uncoupled cell (solid lines) and stimuli $h(t)$ (bars) at
$r=100$ events/sec with $V_0=5$ and $k=0.3$; (b) system with $L=50$ coupled
cells at $r=10$ events/sec: $x_i=1$ (filled circles), $2\leq x_i \leq
n-1$ (open circles) and $x_i=0$ (white background). Arrows indicate
events to be considered in more detail subsequently.}
\end{center}
\end{figure}

In order to visualize the effect of the refractory period, we mimick
the behavior of the spike of a neuron by mapping the automaton state
into an action potential wave form
\begin{eqnarray}
V(x_i) & = & V_0 \left\{ \frac{}{} \delta(x_i,1) -
\left[1-\delta(x_i,0)\right] \left[1-\delta(x_i,1)\right] \right. 
	\nonumber \\
 & \ & \left. \times k\left(1-\frac{(x_i-2)}{n-2}\right) \right\}\; . 
\end{eqnarray}
Notice that $V$ plays no role whatsoever in the
dynamics. Fig.~\ref{fig:dyn}(a) shows the behavior of $V(x_i(t))$ for
an uncoupled 5-state automaton. We observe that stimuli that fall
within the refractory period go undetected, and in the absence of
stimuli the automaton eventually returns to and stays at its quiescent
state $x_i=0$. Since a typical spike lasts the order of 1 ms, this
provides a natural time scale of 1 ms per time step, which will be
used throughout this paper.

\begin{figure}
\begin{center}
\includegraphics[width=0.6\columnwidth, angle=270]{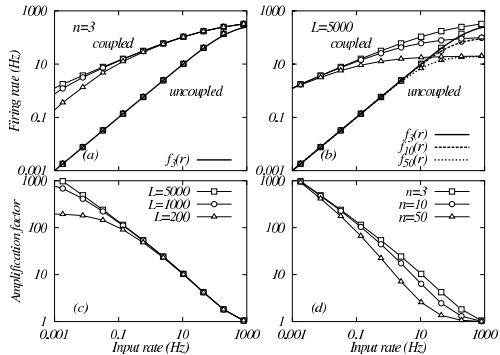}
\caption{\label{fig:loglog} Firing rates $f$ and $F$ (top) and amplification
factor $A$ (bottom) vs. input rate $r$ for $n=3$ and varying $L$ [(a)
and (c)] and for $L=5000$ and varying $n$ [(b) and (d)]. Thick lines (top
panels) show $f_n(r)$ as in Eq.~\ref{f}.}
\end{center}
\end{figure}

Response of uncoupled receptor cells is shown in Fig.~\ref{fig:loglog}
(thick lines on top panels). We draw input signals at rate $r$ per
cell and measure the average firing rate $f$ (spikes per second per
cell) of the $n$-state automata over a sufficiently long time. In the
low rate regime the activity of the uncoupled cells is
proportional to the signal rate.  If the rate increases, there is a
deviation from the linear behavior due to the cell's refractory time
$\Delta_n \equiv n\times 10^{-3}$ seconds. The single-cell response $f$ is
extremely well fitted by a linear-saturating curve $f_n$
[Fig~\ref{fig:loglog}(a) and (b)]:
\begin{equation}
\label{f}
f_{n}(r) =  r/(1+r\Delta_n) \: , 
\end{equation}
which can be deduced from the fact that the firing rate is
proportional to the rate discounting the refractory intervals, $f_n=
r(1-f_n\Delta_n)$. The same result can be obtained by a stationary mean field
solution of the uncoupled cells.

How to improve the sensitivity for very low rates? If we consider the response
$R$ (spikes per second) of the total pool of $L$ independent cells, we have 
$R = Lf \approx  Lr$, so increasing $L$ increases the total sensitivity of the 
epithelium. Although certainly useful, this scaling is trivial since the efficiency
of each cell remains the same. 

Coupled excitable cells (say, via gap junctions) are an example of
excitable media that supports the propagation of nonlinear waves
\cite{Murray}.  Here we show that the formation and annihilation
of these waves enhance the sensitivity and, at the same time, extends
the dynamic range of a sensory epithelium.  We couple $L$ cellular
automata in a chain by defining the local input as
\begin{equation}
h_i(t) = 1-\left[1-\delta(I_{i}(t),1)\right]
\prod_{j=\pm 1}\left[1-\delta(x_{i+j}(t),1)\right]\; ,
\end{equation}
i.e. $h_i(t)$ will be nonzero whenever either of $i$'s neighbors are
spiking and/or the external input is nonzero. This kind of coupling
models electric gap junctions instead of chemical synapses because it
is fast and bidirectional.

A sample of the resulting chain dynamics is shown in
Fig.~\ref{fig:dyn}(b). Due to coupling, single input events create
waves that propagate along the chain, leaving behind a trail of
refractoriness (of width $n-1$) which prevents new spikes from
reappearing immediately.  More importantly, refractoriness is
responsible for wave annihilation: when two wave fronts meet at site
$i$ they get trapped because the neighboring sites have just been
visited and are still in their refractory period. 
This is a well known phenomenon in
excitable media \cite{Murray} and occurs in the CA model $\forall
n\geq 3$. Notice that the overall shape of two consecutive wave-fronts
are correlated (see Fig.~1), denoting some kind of memory effect,
a phenomenon observed previously 
by Chialvo {\em et al.} \cite{Chialvo} and Lewis and Rinzel
\cite{Lewis}. 

Due to a chain-reaction mechanism, the spike of a single receptor cell is
able to excite all the other cells.  The sensitivity per neuron has
thus increased by a factor of $L$. This can be clearly seen in
Fig.~\ref{fig:loglog}, which shows the average firing rate per cell
$F$ in the coupled system (top panels), as well as the amplification
factor $A \equiv F/f$ (bottom panels).  This is a somewhat expected
effect of the coupling: neuron $j$ is excited by signal events
that arrive not only at neuron $j$ but elsewhere in the network. 

More surprising is the fact that the dynamic range (the interval of
rates where the neuron produces appreciable but still non-saturating
response) also increases dramatically.  This occurs due to a second
effect, which we call the self-limited amplification effect.  Remember
that a single spike of some neuron produces a total of $L$ neuronal
responses. This is valid for small rates, where inputs are
isolated in time from each other.  However, for
higher signal rates, a new event occurs at neuron $k$ before the
wave produced by neuron $j$ has disappeared. If the initiation site
$k$ is inside the fronts of the previous wave [e.g. the events
signaled by arrows in Fig.~\ref{fig:dyn}(b)], then two events produce
$2L$ responses as before. But if $k$ is situated outside the fronts of
the $j$-initiated wave [as in the first input events shown in
Fig.~\ref{fig:dyn}(b)], one of its fronts will run toward the $j$-wave
and both fronts will annihilate. 

Thus, two events in the array have produced only $L$ excitations (that
is, an average of $L/2$ per input event). So, in this case, the
efficiency for two consecutive events (within a window defined by the
wave velocity and the size $L$ of the array) has been decreased by half. If
more events (say, $m$) arrive during a time window, many fronts
coexist but the average amplification of these $m$ events (how many
neurons each event excites) is only of order $L/m$.

Therefore, although the amplification for small rates is very high,
saturation is avoided due to the fact that the amplification factor
decreases with the rate in a self-organized non-linear way. The
amplification factor $A$ shown in
Figs.~\ref{fig:loglog}(c)~and~\ref{fig:loglog}(d) decreases in a
sigmoidal way from $A={\cal O}(L)$ for very small rates (since a
single event produces a global wave) to $A=1$ for large rates, where
each cell responds as if isolated since waves have no time to be
created or propagate.

The role of the system size $L$ for low input rates becomes clear in
Fig.~\ref{fig:loglog}(c): the larger the system, the lower the rate
$r$ has to be in order for the amplification factor to saturate at
${\cal O}(L)$. In other words, we can think of a decreasing crossover
value $r_1(L)$ such that the response is well approximated by $F(r)=
Lf(r)\approx Lr$ for $r\ll r_1(L)$. In this linear regime consecutive
events essentially do not interact. Larger system sizes increase not
only the overall rate of wave creation ($\sim 1-(1-r)^L$) 
but also the time it takes for a wave to reach the borders and
disappear.  In the opposite limit of large input rates, the behavior
of the response is controlled by the absolute refractory period
$\Delta$, as shown in Fig.~\ref{fig:loglog}: $F$ and $f$ saturate at
$r_2 \equiv 1/\Delta$ for $f \agt r_2$, independently of the system
size.

\begin{figure}[!tb]
\begin{center}
\includegraphics[width=0.6\columnwidth, angle=270]{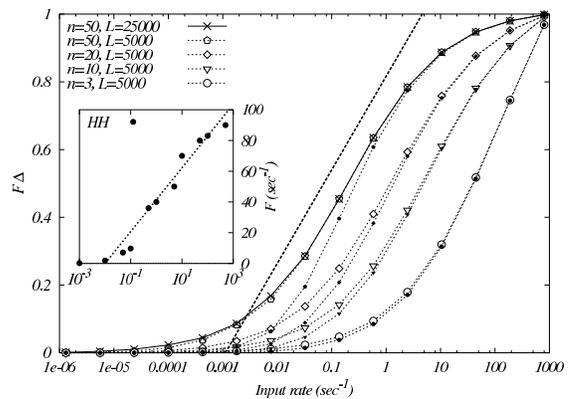}
\caption{\label{fig:semilog}$F\times \Delta$ vs. input rate $r$ for
$L=5000$ (open symbols) and $L=200$ (filled symbols) for different
values of $n$. A $L=25000$ curve for $n=50$ (crosses)
shows no difference to the $L=5000$ case.
Straight lines are intended as a guide to
the eye. Inset: $F(r)$ for the Hodgkin-Huxley system.}
\end{center}
\end{figure}

So what happens for intermediate input rates, i.e. $r_1\ll r \alt
r_2$? The answer is a slow, Weber-Fechner-like increase in the
response $F$, as can be seen in Fig.~\ref{fig:semilog}. The
logarithmic dependence on $r$ is a good fit of the curves for about
three decades. 

Motivated by results obtained with more realistic 
elements \cite{Kinouchi} we introduced a relative refractory
period in our CA model. We first define a time window $M$ after a
spike during which no further spikes can occur (absolute refractory
period). In the following $n-M-2$ steps (relative refractory period), a
single input does not produce a spike but two or more inputs can
elicit a cell spike if they arrive within a temporal summation window
$\tau$ (details of this model will be described in a forthcoming
full paper). This ingredient produced the appearance of a power law
$F(r)$ curve (Stevens Law \cite{Stevens,BBS}), as shown in
Fig.~4. Notice that the exponent depends on the relative refractory
period. The appearence of a power law transfer function
is a robust effect also observed in
coupled maps systems \cite{Kinouchi}.

\begin{figure}[!tb]
\begin{center}
\includegraphics[width=0.6\columnwidth, angle=270]{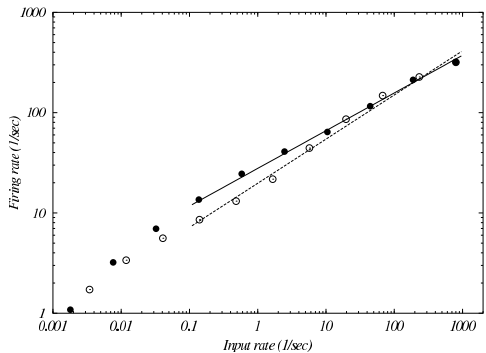}
\caption{\label{fig:stevens} Neuronal ``Stevens Law'' $F \propto
r^\alpha$ in automata which takes temporal summation effects into
account (see text for details). Firing rate $F$ vs. input rate $r$ for
a CA with $n$ states and an absolute refractory period of $M=3$ time
steps. Filled circles: $n=15$, $\tau=10$, $\alpha = 0.38$; open
circles: $n=100$, $\tau=80$, $\alpha=0.44$.}
\end{center}
\end{figure}

\begin{figure}[t]
\begin{center}
\includegraphics[width=0.6\columnwidth, angle=270]{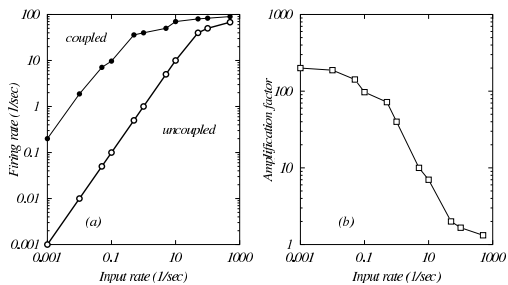}
\caption{\label{fig:HH} (a) Firing rate for coupled ($F$, filled circles)
and uncoupled ($f$, open circles) systems and (b) amplification factor $A=F/f$
vs. input rate $r$ for Hodgkin-Hukley neurons for $L=200$.}
\end{center}
\end{figure}

We may confirm the generic character of the self-regulated
amplification phenomenon by performing simulations using biophysically
detailed cell models, for example a network of Hodgkin-Huxley (HH)
elements with the standard set of parameters given in \cite{Genesis}
connected via gap junctions of $30 M\Omega$.  Preliminary results
show that this system exbits the same qualitative behavior of the simple CA model 
(see inset of Fig.~\ref{fig:semilog} and Fig.~\ref{fig:HH}). 
More detailed results will be reported elsewhere.

Concerning the functional role of gap junctions for signal processing,
it has been recognized that they provide faster communication between
cells than chemical synapses and play a role in the synchronization of
cell populations \cite{Traub}. Here we are proposing another
functional role for gap junctions: the enhancement of the dynamic
range of neural networks.

There is considerable debate about what is the most appropriate
functional law to describe psycophysical response: Weber-Fechner,
Stevens or some interpolation between the two \cite{BBS}. Our results
suggest that properties of excitable media could be a bottom-up
mechanism which can generate both laws, and a cross-over between them,
depending on the presence of
secondary factors like the relative refractory periods and temporal
summation.

We can even make two more specific predictions
which are easily testable experimentally: 
1. The larger the relative refractory period (e.g., due to slower
hyperpolarizing currents) of sensory epithelia neurons, the larger the
exponent of Stevens Law; 2.
For sufficiently low input rates, the sensory epithelium response will
be always linear ($\alpha=1$).

This mechanism for amplified but self-limited response due to
wave annihilation promotes signal compression,
is a basic property of excitable media 
and is not restricted to one dimensional systems. We
conjecture that the same mechanism could be implemented at different
biological levels, from hippocampal networks (where axo-axonal 
gap junctions have been recently reported \cite{Traub} and modeled \cite{Lewis}
by a CA similar to ours) to excitable dendritic trees in single
neurons \cite{Chialvo,Koch}.  This signal compression mechanism could
also be implemented in artificial sensors based on excitable media.

{\bf Acknowledgments:} Research supported by FAPESP, CNPq and
FAPERJ. The authors thank Silvia M. Kuva for valuable suggestions and
the referees for useful comments. OK thanks the hospitality of
Prof. David Sherrington and the Theoretical Physics Department of
Oxford University where these ideas have been first developed.

\end{document}